\documentclass[conference]{IEEEtran}
\usepackage[hidelinks,pdfauthor={Alvari Kupari; Nasser Giacaman; Valerio Terragni},
            pdftitle={Adoption and Evolution of Code Style and Best Programming Practices in Open-Source Projects},
            pdfkeywords={Software quality, Code style conventions, Coding standards, Open-source repositories, Software development practices, Software metrics},
            pdfsubject={This paper presents a large-scale empirical study of 1,036 popular open-source Java repositories to examine the adoption and evolution of code style conventions and best programming practices.}]{hyperref}
\usepackage{flushend}
\usepackage{amsmath,amssymb,amsfonts}
\usepackage{algorithmic}
\usepackage{eso-pic}

\usepackage{graphicx}
\usepackage{textcomp}
\usepackage{xcolor}
\newcommand{\checkstyle}{\textsc{\mbox{CheckStyle}}\@\xspace}
\usepackage{xspace}
\newcommand{\toolname}{GradeStyle}
\newcommand{\autotool}{\textsc{\toolname}\@\xspace}
\newcommand{\tool}{\autotool}
\usepackage[table]{xcolor}

\usepackage{listings}
\usepackage{xcolor}
\newcommand{\category}[1]{\smallskip
\par
\noindent
\textbf{#1}}

\usepackage{xcolor}
\usepackage{ragged2e} % Provides the \justifying command

\usepackage{cite}
\usepackage{tcolorbox}
\usepackage{ifthen}
\usepackage{graphicx}
\usepackage{tcolorbox}
\usepackage{booktabs}

\newtcolorbox{custombox}[1]{
	colback=gray!10,
	colframe=gray!70,
	left=1.5mm,
	right=1.5mm,
	top=1.5mm,
	bottom=1.5mm,
	fonttitle=\bfseries,
	arc=0mm,
	leftrule=1mm,
	rightrule=0mm,
	toprule=0mm,
	bottomrule=0mm,
	notitle,
	before=\par\medskip\medskip\noindent,
	before upper={\textbf{#1: } },
}

\definecolor{codeblue}{rgb}{0.22, 0.45, 0.70}
\definecolor{codegreen}{rgb}{0.0, 0.5, 0.0}
\definecolor{codered}{rgb}{0.6, 0, 0}
\definecolor{codegray}{rgb}{0.5, 0.5, 0.5}

\lstdefinestyle{java}{
  language=Java,
  basicstyle=\ttfamily\footnotesize,
  keywordstyle=\color{codeblue}\bfseries,
  stringstyle=\color{codered},
commentstyle=\color{codegreen}\itshape,
  numbers=left,
  numberstyle=\tiny\color{codegray},
  stepnumber=1,
  numbersep=10pt,
  backgroundcolor=\color{white},
  showspaces=false,
  showstringspaces=false,
  tabsize=4,
  breaklines=true,
  breakatwhitespace=true,
  frame=single,
  rulecolor=\color{codegray},
  captionpos=b,
  escapeinside={(*@}{@*)},
      abovecaptionskip=10pt
}

\lstdefinestyle{console}{
basicstyle=\ttfamily\footnotesize, % Set text to white
    backgroundcolor=\color{gray!20}, % Set background to black
    frame=single,
        frame=none, % Remove the frame (box line) around the code
    breaklines=true,
    keepspaces=true,
    columns=fixed,
    deletekeywords={new, for},
    abovecaptionskip=10pt,
        numbers=none % Disable line numbers
}

\usepackage{float}
\usepackage{placeins}
\usepackage{fancybox,framed}
\usepackage[utf8]{inputenc}

\usepackage[figure,norelsize,vlined,linesnumbered]{algorithm2e}
\SetAlFnt{\small}
\SetAlCapFnt{\large}
\SetAlCapNameFnt{\large}

\SetCommentSty{mycommfont}

\usepackage{booktabs} % For formal tables
\usepackage{xspace}
\usepackage{tabularx,colortbl}
\usepackage{multirow, makecell}
\usepackage{paralist}
\usepackage{tabularx}
\usepackage{listings}
\usepackage{graphicx}
\usepackage{ifthen}
\usepackage{amsthm}
\usepackage{indentfirst}
\usepackage{xcolor}

\usepackage{tikz}
\usetikzlibrary{backgrounds}
\usepackage{booktabs}

\usepackage{pifont}
\definecolor{darkjunglegreen}{rgb}{0.000000,0.392157,0.000000}

\usepackage{url}
\def\codesize{\small}%\smaller
\def\<#1>{\codeid{#1}}\protected\def\codeid#1{\ifmmode{\mbox{\codesize\ttfamily
			#1}}\else{\codesize\ttfamily
		#1}\fi}

\definecolor{mygray}{gray}{0.9}
\usepackage{fancybox}
\usepackage{listings}
\usepackage{etoolbox}
\usepackage{kvoptions}
\makeatletter
\newcommand{\removelatexerror}{\let\@latex@error\@gobble}
 
\usepackage{xspace}
\newcommand{\java }{\textsc{Java}\xspace}
\newcommand{\gh }{\textsc{GitHub}\xspace}

\usepackage{csvsimple}

\SetKwProg{Fn}{Function}{}{}
\makeatother

\usepackage{booktabs} 
\usepackage{graphics} % for resizebox tables
\usepackage{multirow,bigstrut}
\usepackage[framemethod=tikz]{mdframed}
\usepackage{graphicx}
\usepackage{bm} %bold math
\usepackage{relsize}

\usepackage{bold-extra}

\DeclareMathAlphabet\mathbfcal{OMS}{cmsy}{b}{n}

\usepackage{multirow,array,varwidth}
\usepackage{booktabs}
\usepackage{pifont}

\usepackage{listings}
\usepackage{color,url}

\definecolor{javared}{rgb}{0.6,0,0} % for strings
\definecolor{javagreen}{rgb}{0.25,0.5,0.35} % comments
\definecolor{javapurple}{rgb}{0.5,0,0.35} % keywords
\definecolor{javadocblue}{rgb}{0.25,0.35,0.75} % javadoc
\usepackage{lipsum}
\usepackage{paralist}

\newtheoremstyle{mystyle}%                % Name
{0.0mm}%                                     % Space above
{0.0mm}%                                     % Space below
{}%                             % Body font \itshape
{}%                                     % Indent amount
{\bfseries}%                            % Theorem head font
{.}%                                    % Punctuation after theorem head
{ }%                                    % Space after theorem head, ' ', or \newline
{}%                                     % Theorem head spec (can be left empty, meaning `normal')

\usepackage{amsbsy}
\definecolor{orange}{rgb}{1,0.5,0}
\definecolor{darkjunglegreen}{rgb}{0.000000,0.392157,0.000000}

\usepackage{soul}

\usepackage{paralist}

\font\tt=rm-lmtl10

\usepackage{enumitem}
\usepackage{amsmath}
\usepackage{comment}
\usepackage{microtype}
\usepackage{amsthm}

\newlength\marincrease



\begin{document}

\title{Adoption and Evolution of Code Style and \\ Best Programming Practices in Open-Source Projects}

%\author{\IEEEauthorblockN{Anonymous Author(s)}}

\author{\IEEEauthorblockN{Alvari Kupari}
\IEEEauthorblockA{University of Auckland\\
Auckland, New Zealand \\
akup390@aucklanduni.ac.nz
}
\and
\IEEEauthorblockN{Nasser Giacaman}
\IEEEauthorblockA{University of Auckland\\
Auckland, New Zealand \\
n.giacaman@auckland.ac.nz \\
}
\and
\IEEEauthorblockN{Valerio Terragni}
\IEEEauthorblockA{University of Auckland\\
Auckland, New Zealand \\
v.terragni@auckland.ac.nz \\
}
}

\maketitle
\AddToShipoutPictureFG*{%
  \AtPageLowerLeft{%
    \raisebox{1.6cm}{%
      \hspace*{3.0cm}%
      \parbox{\dimexpr\paperwidth-5.0cm\relax}{
      This is the authors’ version of the paper published in IEEE International Conference on
      Software Maintenance and Evolution (ICSME 2025).
      DOI: \href{https://doi.org/10.1109/ICSME64153.2025.00026}{10.1109/ICSME64153.2025.00026}.}%
    }%
  }%
}

	\begin{abstract}

Following code style conventions in software projects is essential for maintaining overall code quality. Adhering to these conventions improves maintainability, understandability, and extensibility. Additionally, following best practices during software development enhances performance and reduces the likelihood of errors. This paper analyzes 1,036 popular open-source \java projects on \gh to study how code style and programming practices are adopted and evolve over time, examining their prevalence and the most common violations. Additionally, we study a subset of active repositories on a monthly basis to track changes in adherence to coding standards over time. We found widespread violations across repositories, with Javadoc and Naming violations being the most common. We also found a significant number of violations of the \textsc{Google} Java Style Guide in categories often missed by modern static analysis tools. %, such as  such as \emph{Missing Override} and \emph{Unqualified Static Access}. 
Furthermore, repositories claiming to follow code-style practices exhibited slightly higher overall adherence to code-style and best-practices. The results provide valuable insights into the adoption of code style and programming practices, highlighting key areas for improvement in the open-source development community. Furthermore, the paper identifies important lessons learned and suggests future directions for improving code quality in \java projects.
\end{abstract}


\section{Introduction}

Following code style conventions is essential for maintaining high-quality software~\cite{allamanis2014learning}. Proper code style improves readability, maintainability, and consistency, making it easier for developers to collaborate and extend software projects~\cite{tornhill2022codered}. Given the prominent role of open-source software~\cite{jonsson2023opensource}, understanding the degree to which projects adhere to established style conventions is increasingly important.

\smallskip
Previous studies on code style and programming practice adoption~\cite{elish2002adherence, Beller2016, Boogerd2008, brown2022novice, jonsson2023opensource,butler2011mining} have been limited in scale. Many focused on small datasets and small number of code-style checkers. Additionally, it remains unclear how adherence to code-style evolves over time or whether explicitly claiming adherence in documentation leads to better code-style.

\smallskip
To address these gaps, this paper presents a large-scale empirical study on \java code style and best programming practice adherence. Specifically, we rely on \textsc{Google}’s \java Style Guide~\cite{google}, which is widely considered the most popular \java coding convention.
To conduct this study, we extended \tool~\cite{gradestyle}, an open-source tool we previously developed as a style checker for assessing the code style of student assignments.  Our extension allows the tool to analyze open-source repositories. We also implemented eight new code style and programming practice checkers based on \textsc{Google}’s \java Style Guide to improve the tool’s ability to assess real-world projects. To the best of our knowledge, the extended version of \tool provides the most comprehensive detection of \textsc{Google} \java code style violations available. Indeed, commercial tools like PMD and \checkstyle miss important violations~\cite{gradestyle}.

\smallskip
We analyzed 1,036 open-source \java repositories from \gh, selected based on their popularity (ranked by stars). By running \tool on these repositories, we studied the adoption of nine code style and seven programming practices, identifying key areas for improvement. Additionally, we distinguished between repositories that explicitly claim to follow a style guide and those that do not, allowing us to examine whether stated intent correlates with actual adherence. Moreover, from this dataset, we selected 41 mature and actively maintained repositories for further analysis. Each repository was analyzed 12 times—once per calendar month over the past year. This allowed us to track how adherence to coding standards evolves over time.

\begin{comment}
\smallskip
In particular, this study addresses the following key research questions:
\begin{itemize}
\item \textbf{RQ1:} How prevalent are different code style and programming practice violations in open-source \java projects?
\item \textbf{RQ2:} Do repositories that explicitly claim to follow a code style guide exhibit better adherence to coding standards than those that do not?
\item \textbf{RQ3:} How does adherence to code style and programming practices evolve over time in mature and actively maintained \java repositories?
\end{itemize}
\end{comment}

\smallskip
The results of our study reveal several key findings: 
First, code-style and best-practice violations are widespread, with naming conventions and Javadoc documentation being among the most frequently violated categories.
Second, repositories that explicitly claim to follow a coding standard exhibit slightly better adherence.
Third, over time, adherence to programming best practices shows modest improvement, while certain code style violations, such as method and variable naming, become more prevalent.

\smallskip
In summary, this paper makes the following contributions:
\begin{itemize}
\item We present a large-scale empirical study of 1,036 popular open-source \java repositories, providing a comprehensive overview of code style and best programming practice adherence.
\item We bring a series of important insights into the adoption and evolution of code style and best programming practice adherence. Our findings provide actionable insights for open-source maintainers, tool developers, and researchers seeking to improve coding standards in \java projects.
\item We propose a new version of \tool for analyzing open-source repositories, with eight new violation checks based on \textsc{Google}’s \java Style Guide. We released the new version of the tool~\cite{gradestyletool} and all experimental data~\cite{data} to foster future work in this area.
\end{itemize}

\section{Experimental Design}
Our study investigates the adoption and evolution of code style, programming practices in popular open-source \java projects in \gh. Specifically, our empirical study addresses the following four research questions (RQs):

\begin{itemize}
\item[\textbf{RQ1}]\textbf{Violation Prevalence.} How prevalent are different code style and programming practice violations in open-source \java projects?
\item[\textbf{RQ2}] \textbf{Claimed vs. Actual Code Style Adherence.} Do repositories that explicitly claim to follow a code style guide (e.g., \textsc{Google}’s \java Style Guide) exhibit better adherence to coding standards than those that do not?
\item[\textbf{RQ3}] \textbf{Evolution of Code Style and Best Programming
Practices.} How does code style and programming practice adherence evolve over time in actively maintained and mature \java repositories?
\item[\textbf{RQ4}] \textbf{Common Orderings of Class Elements.} Is there a common ordering of \java elements in a class that is widespread in open-source repositories?
\end{itemize}

\smallskip
\textbf{RQ1} aims to understanding the extent to which code style and programming practice violations occur, providing insights into common issues that developers face. Identifying frequent violations can highlight areas where better enforcement or improved tooling may be needed.

\smallskip
\textbf{RQ2} investigates whether explicit mentions of code style in documentation or configuration files correlate with better compliance, providing insights into the effectiveness of declaring coding style adherence.

\smallskip
\textbf{RQ3} analyzes the historical data (i.e., git commits) of well-maintained and actively developed \java repositories to assess how adherence to code style and programming practices evolves over time. This investigation helps determine whether long-term maintenance leads to improvements in best practices and overall code quality.

\smallskip
\textbf{RQ4} aims to understand the prevalence of common orderings of \java elements within a class.  While \java style guides, such as \textsc{Google}’s, recommend consistency, they do not prescribe a specific order. As a result, there is no universally accepted standard. Identifying popular orderings can help developers identify and follow commonly used, but not strictly enforced standards for ordering Class elements.

\subsection{Code Style Violations and Best Practices}
\label{violations}
We now explain the code style violations and programming best practices analyzed in this study. The two most widely known coding conventions for \java are \textsc{Oracle}'s (Sun's) \java Code Conventions~\cite{oracle} and \textsc{Google}'s \java Style Guide~\cite{google}. These conventions are mostly similar, with minor differences. We selected \textbf{\textsc{Google}'s \java Style Guide} because \textsc{Oracle}'s conventions have not been updated since April 20, 1999 (over 23 years ago). \textsc{Google}'s guide, last updated in 2018, covers modern \java features introduced after 1999 and reflects how \java code style has changed over time~\cite{trautsch2020longitudinal}.

\smallskip
We did not consider every possible violation or programming practice specified in the \textsc{Google}'s \java Style Guide~\cite{google}, as some of them vary depending on tools and developer preferences. For example, indentation, brace placement, and whitespace formatting usually follow \textsc{Google}'s \java Style Guide, but formatting tools in IDEs may produce slightly different results~\cite{piantadosi2020does}. These minor differences do not significantly impact code quality, so we excluded them from our study. Another example is the order of elements within a class. Although the order is crucial for code readability, as emphasized in \textsc{Google}'s \java Style Guide~\cite{google}, \textsc{Google}'s guide does not enforce a specific order. Thus, we did not include ordering as a violation but still analyzed in RQ4 different ordering patterns to find common practices. Moreover, we incorporated additional checkers for \java best programming practices violations that we consider important, common, and interesting to analyze.

\smallskip
Table~\ref{tab:violations-tools} summarizes the violations analyzed in this study, categorized into two groups according to the \textsc{Google} \java Style Guide. 

\begin{itemize}
\item \textbf{Code style violations}: These violations focus on how code is written, particularly regarding naming conventions and formatting standards. Adhering to these guidelines enhances readability and maintainability, making future modifications easier. 
\item \textbf{\java best programming practices violations}: Unlike general code style, these violations relate specifically to \java coding practices. Although following these practices can also improve readability, their primary goal is to enhance maintainability, security, and performance.
\end{itemize}

\medskip
\noindent \textbf{Code style violations:}

\category{Class Names.}
Class and enum names must be in \textit{UpperCamelCase} (each internal word starts with a capital letter). Class names should be nouns or noun phrases, such as \texttt{Customer} or \texttt{DataManager}. \textsc{Google}'s \java Style Guide allows class names to start with adjectives followed by nouns, like \texttt{SortedMap} or \texttt{ImmutableSet}. Any class declaration not following these naming rules counts as a violation.

\category{Package Names.}
Package names must be lowercase and match the directory structure. For example, a class \texttt{User.java} located in \texttt{src/org/project/User.java} must declare the package as \texttt{package org.project;}. Package names violating this rule count as violations.

\category{Method Names.}
Method names should be verb phrases using \textit{lowerCamelCase} (the first word starts lowercase, with each following word capitalized). Method declarations that break this rule count as violations.

\category{Variable Names.}
Variable names, including instance fields, parameters, and local variables, should use \textit{lowerCamelCase} and not start with an underscore (\_) or dollar sign (\$). Static final variables must use uppercase letters with underscores separating words (e.g., \texttt{TOTAL\_COUNT}, \texttt{COLOR\_BLUE}). Variable names breaking these conventions count as violations.

\category{Javadoc.}
Every public class, method, enum, and constructor must have a Javadoc comment, and these comments should contain a minimum number of words. While \textsc{Google}'s \java Style Guide does not explicitly mention how long Javadocs should be, we required a minimum of 10 words to ensure clarity and depth in Javadoc comments. This promotes detailed documentation across projects. Method Javadocs must also include appropriate tags like \texttt{@param}, \texttt{@return}, and \texttt{@throws}, matching their method signatures. Specifically:
\begin{itemize}
\item \textbf{Classes and Enums}: Must have a Javadoc comment above the declaration, with at least 10 words.
\item \textbf{Constructors}: Must have a Javadoc comment above the declaration, with at least 10 words.
\item \textbf{Methods}: Must have a Javadoc comment above the declaration, with at least 10 words.
\end{itemize}

\medskip
\noindent \textbf{\java best programming practices violations:}

\category{Missing Override.}
Methods overriding or implementing superclass or interface methods must include the \texttt{@Override} annotation, unless the parent method is marked as \texttt{@Deprecated}. This annotation prevents errors, improves readability, and ensures refactoring safety.

\category{Empty Catch Blocks.}
Caught exceptions should never be silently ignored. Typical actions include logging the error or rethrowing an exception. If intentionally left empty, the reason must be clearly explained with a comment:

\begin{lstlisting}[language=Java]
try {
  int value = Integer.parseInt(input);
  processNumber(value);
} catch (NumberFormatException ok) {
  // Non-numeric input is expected; continue normally
}
\end{lstlisting}

In test methods, exceptions with names starting with \texttt{expected} may be ignored without a comment, as shown below:

\begin{lstlisting}[language=Java]
try {
  emptyStack.pop();
fail();
} catch (NoSuchElementException expected) {}
\end{lstlisting}

\category{Unqualified Static Access.}
Static class members must be accessed using the class name, not through an instance or method return value:

\begin{lstlisting}[language=Java]
// doWork() is a static method
Utils.doWork(); // good
utilInstance.doWork(); // bad
getUtils().doWork(); // very bad
\end{lstlisting}

\category{Finalize Override.}
Methods should never override \texttt{Object.finalize()}, since \java has scheduled finalization for removal due to reliability issues.

\category{Private Instances.}
To support encapsulation, instance fields should be declared \texttt{private} or \texttt{protected}. External access to these fields should use public getter and setter methods.

\category{String Concatenation.}
Frequent string concatenation, especially in loops, reduces performance significantly. The following pattern should be avoided:

\begin{lstlisting}[language=Java]
String result = "";
for (int i = 0; i < 50000; i++) {
   result += i + " ";
}
\end{lstlisting}

Instead, use a \texttt{StringBuilder}:

\begin{lstlisting}[language=Java]
StringBuilder sb = new StringBuilder();
for (int i = 0; i < 50000; i++) {
   sb.append(i).append(" ");
}
String result = sb.toString();
\end{lstlisting}

The first example can be over 500 times slower than the second\footnote{While the Java Just-In-Time (JIT) compiler may optimize simple string concatenations using \texttt{StringBuilder}, this optimization often does not apply to repeated concatenation in loops. Each iteration creates a new \texttt{String} object, leading to significant performance overhead due to the immutability of \texttt{String}. Using \texttt{StringBuilder} avoids this by reusing a mutable buffer.}. Each instance of string concatenation in loops counts as a violation.

\category{Useless Code.}
High-quality code should not contain unused imports, methods, variables, or commented-out code. Every line containing such unused or commented-out code counts as a violation.

\subsection{Implementation}
To detect code style violations, we evaluated several existing tools, initially focusing on well-known commercial tools such as \checkstyle~\cite{checkstyle}, PMD~\cite{pmd}, \textsc{SonarQube}\cite{sonarqube}, \textsc{ErrorProne}\cite{errorprone}, and \textsc{SpotBugs}~\cite{spotbugs}. However, these tools do not aim to detect many violations described in \textsc{Google}'s \java Style Guide. For instance, although these tools can detect class and method naming conventions (\textit{UpperCamelCase} and \textit{lowerCamelCase}), they do not apply natural language processing (NLP) to ensure correct noun or verb usage, which is required by \textsc{Google}'s \java Style Guide. Indeed, \checkstyle marks \textsc{Google}'s rules ``5.2.2-Class-names'' and ``5.2.3-Method-names'' as partially supported~\cite{checkstyleGoogle}.

\smallskip
These existing tools also struggle detecting violations which require context specific information. For example, \checkstyle cannot detect violations requiring context from multiple files simultaneously, such as \textit{Missing Override} and \textit{Unqualified Static Access}. This is because \checkstyle analyses one file at a time, and is therefore unable to deduce the class hierarchy needed to detect such violations. Indeed, \textsc{Google}'s \java style guide rules 6.1 and 6.3 are explicitly marked as ``cannot be detected by Checkstyle due to its limitation: multiple file checking is not supported''~\cite{checkstyleGoogle}.

\begin{table*}[t]
\rowcolors{1}{}{gray!10}

    \centering
    \caption{Enabled Violation Checkers (\ding{80} indicates newly implemented violations not originally supported by \tool~\cite{gradestyle})}
    \resizebox{1\linewidth}{!}{%
    \renewcommand{\arraystretch}{1}
    \setlength{\tabcolsep}{2pt}
    \begin{tabular}{lllccc}
    \hiderowcolors
        \toprule
        \textbf{Violation} & \textbf{Description} & \textbf{Impl.} & \textbf{\tool} & \textbf{\textsc{Google}} & \textbf{Novel}  \\
        \midrule
        \multicolumn{5}{c}{\textbf{Code Style Violations}} \\
        \hline
        \showrowcolors
        Class Names & Class names should have a noun, follow UpperCamelCase. & custom  & \checkmark & \checkmark & \ding{55} \\
        Method Names & Method names should have a verb, follow camelCase. & custom & \checkmark & \checkmark & \ding{55} \\
         Variable Names & Variable names should follow camelCase. & custom  & \checkmark & \checkmark & \ding{55} \\
         Package Names & Package names should match the project structure and be all lowercase. & \textsc{CS}~\cite{checkstyle}  & \checkmark & \checkmark & \ding{55} \\
         Javadoc Formatting & Javadoc comments should be there, with comments for all tags. & \textsc{CS}~\cite{checkstyle} & \checkmark & \checkmark & \ding{55} \\
         Class Javadocs & Classes should have a Javadoc comment above the class declaration. & custom & \ding{80} & \checkmark & \ding{55} \\
         Constructor Javadocs & Constructors should have a Javadoc comment above the constructor declaration (min 10 words). & custom & \ding{80} & \checkmark & \ding{55} \\
         Method Javadocs & Methods should have a Javadoc comment above the method declaration (min 10 words). & custom & \ding{80} & \checkmark & \ding{55} \\
         Field Javadocs & Fields should have a Javadoc comment. & custom & \ding{80} & \checkmark & \ding{55} \\
             \hiderowcolors

        \hline
        \multicolumn{5}{c}{\textbf{Programming Practices Violations}} \\
        \hline
        \showrowcolors
         Private Instances & All fields within a class should be private to prevent outside access; getters/setters should be used. & custom & \checkmark & \ding{55} & \ding{55} \\
         Useless & For exmaple, fields and methods that are not used are flagged as unnecessary. & custom & \checkmark & \ding{55} & \ding{55} \\
         String Concatenation & Strings should not be concatenated within loops to boost performance; \texttt{StringBuilder} should be used. & custom & \checkmark & \ding{55} & \checkmark \\
         Missing Override & \texttt{@Override} annotations should be placed wherever a method is overridden for readability. & PMD~\cite{pmd} & \ding{80} & \checkmark & \checkmark \\
         Empty Catch Block & Empty catch blocks can hide errors and make debugging difficult. & PMD~\cite{pmd} & \ding{80} & \checkmark & \ding{55} \\
         Unqualified Static Access & Static methods should be called using the class name to avoid confusion. & custom & \ding{80} & \checkmark & \checkmark \\
         Finalize Override & \texttt{Object.finalize()} should not be overridden, as it is scheduled for removal. & custom & \ding{80} & \checkmark & \checkmark \\
        \bottomrule
    \end{tabular}
    }
    \label{tab:violations-tools}
\end{table*}

\smallskip
We then turned our attention to research tools and selected \textbf{\tool}~\cite{gradestyle}, our own open-source tool, originally developed for educational purposes~\cite{gradestyle}. \tool leverages existing static analysis libraries like PMD and \checkstyle but proposes additional checks using static analysis and NLP techniques. This tool can detect 12 significant categories of \java code style and best-practice violations~\cite{gradestyle}. 
To the best of our knowledge, 	\tool is the most comprehensive tool currently available for detecting \textsc{Google} \java style violations\footnote{Although \checkstyle appears to support many rules from the \textsc{Google} \java Style Guide (see~\cite{checkstyleGoogle}), this is misleading. The site lists sub-rules individually, creating the impression of broader coverage. Conversely, \tool groups related checks together.}.

\smallskip
While \tool already handles several violations targeted in our study, it misses some key checks listed in \textsc{Google}'s \java best practices~\cite{google}, such as \textit{Missing Override} and \textit{Empty Catch Block}. \tool is also built for an educational context where student repositories follow a predictable skeleton structure given by instructors, and is less suitable for open-source analysis where file structure varies considerably. Therefore, we extended \tool to more comprehensively cover these missed checks, and to improve ease of use for research. 

\smallskip 
We also improved \tool's granularity in detecting Javadoc violations. Originally, \tool had a single configuration for Javadocs~\cite{gradestyle}. We modified \tool to separately evaluate constructor, method, field, and class Javadocs. This improvement allows a more detailed and accurate analysis of Javadoc compliance. These additional violation checkers are crucial because adhering to them significantly enhances code readability and understandability, particularly valuable in open-source projects where many contributors collaborate without centralized oversight. 

\smallskip
However, certain original \tool violation checkers were intentionally excluded from our analysis due to limited relevance for open-source projects or because the   results depend heavily on parameter choices, which are difficult to set universally for all projects (e.g., amount of comments~\cite{gradestyle}). 

\smallskip
Table~\ref{tab:violations-tools} summarizes each violation analyzed in this study (see Section~\ref{violations}), indicating whether \tool originally supported it or if we implemented it as a new extension in \tool (marked with \ding{80}). It also indicates which violations are specified in the \textsc{Google} \java Style Guide and which are not (Column ``\textsc{Google}''). We also indicate which violation checkers are novel and, to the best of our knowledge, has not been investigated in open-source repositories before (e.g.,~\cite{elish2002adherence,jonsson2023opensource}), by marking them in the ``Novel'' column.

\smallskip
The original \tool integrates directly with \textsc{Checkstyle}\cite{checkstyle} (v. 10.3.2) for \emph{PackageNames} and \emph{Javadoc} violations. For detecting naming violations (\emph{ClassNames} and \emph{MethodNames}), \tool uses regular expressions combined with NLP through \textsc{extJWNL}~\cite{extjwln} (v. 2.0.5). Regular expressions check for correct camel-case notation, while \textsc{extJWNL} ensures class names contain appropriate nouns and method names contain appropriate verbs by performing Part-of-Speech (PoS) tagging. For other violations \tool implements custom static analyses using \textsc{JavaParser}~\cite{javaparser} (v. 3.23.1), a library designed for static analysis of \java source code, which helps to precisely identify specific language constructs.

\smallskip
To add new violation checkers to \tool, we relied on \textsc{PMD}~\cite{pmd} (v. 6.52.0) for detecting the programming practice violations \texttt{@Override} annotations and empty catch blocks. For the other new violation checkers, we implemented custom static analyzers using \textsc{JavaParser}.

\smallskip
Furthermore, we adapted \tool to better suit open-source projects. Originally, \tool searched for \java files exclusively within the \texttt{src/main/java} directory at the project's root, a standard setup for student assignments. Open-source repositories, however, often follow different and non-standard directory structures (e.g., with \textsc{Maven} submodules). Therefore, we modified the tool to recursively search for \java files across multiple \texttt{src/main/java} directories located anywhere within the project's hierarchy.

\subsection{Repository Mining}
In this study, we aimed to collect a representative set of \textbf{high-quality \java repositories from \gh} by selecting projects in descending order based on their number of stars. Following previous research on mining repositories~\cite{kalliamvakou2014promises}, the number of stars is widely regarded as the best proxy for popularity and, consequently, quality~\cite{kalliamvakou2014promises}. Popular repositories are expected to have higher code quality, a larger number of contributors, and better overall maintenance~\cite{borges2018s}. To balance generalizability with computational efficiency, we set a threshold of \textbf{1,036 repositories}. We retrieved the top-starred \java repositories using the \gh API~\cite{hub4j2025} and stopped after obtaining 1,036 valid repositories.

\smallskip
To ensure meaningful analysis, a valid repository had to meet three selection criteria:

\medskip
\begin{enumerate}
\item \textbf{Repository Size}  
Repositories larger than 2 GB were excluded to avoid excessive processing overhead.

\item \textbf{\java Codebase Size}  
We also excluded repositories with excessively large \java codebases. The \gh API provides a heuristic estimate of total bytes written in Java, which we used as an approximation of the codebase size. Repositories exceeding 2 million bytes of \java code were filtered out to ensure smooth execution of our analysis.

\item \textbf{Tool Compatibility }To verify compatibility, we initially conducted a preliminary check using \checkstyle. Repositories containing \java files with syntax errors that caused tool failures were excluded from the dataset. We chose \checkstyle for this initial check because it is computationally less expensive compared to \tool. If a repository successfully passed the \checkstyle test, we then ran our modified version of \tool to ensure it operated without crashes.
\end{enumerate}

\smallskip
The final dataset included 1,036 repositories with the following characteristics:

\begin{itemize}
\item \textbf{Stars} – Average: 4,816; Min: 1,814; Max: 49,470; Median: 3,098
\item \textbf{Lines of Code (LOC)} – Average: 19,672; Min: 29; Max: 8,873,014; Median: 5,364
\end{itemize}

\smallskip
Due to the filtering criteria, our dataset does not consist solely of the top 1,036 \gh repositories ranked by stars. Instead, it includes the first 1,036 repositories that met all selection criteria. Despite this, all selected repositories have at least 1,814 stars, indicating significant popularity. Therefore, our dataset remains representative of popular open-source \java projects. We then ran our extended version of \tool for all the 1,036 repositories.

\subsection{Manual Validation of Violation Detection}

To validate the accuracy of \tool's violation detection and to quantify the amount of false positives, the first author  performed a manual validation of the results using stratified random sampling. Indeed, manually verifying all detected violations would not be possible. Stratification ensures a more representative evaluation across different violation categories~\cite{singh1996stratified}. We determined the required sample size using a standard calculator~\cite{samplesize2025} with a 15\% margin of error and a 90\% confidence level, resulting in a sample size of 31 violations per category.

\smallskip
For stratification, we divided the 1,036 repositories into 31 evenly sized groups. From each group, we randomly selected one violation per category. We selected the first violation for the category found within the group. 

\smallskip
\tool provides direct links to flagged violation locations, facilitating the verification process. To verify a violation, the first author opened the file and line location specified by the violation report, and then performed a visual inspection to confirm both the presence of the violation, and whether the violation description accurately reflected the code.

\smallskip
Table~\ref{tab:violation-checking} summarizes the number of violations manually checked for each category. The manual validation confirmed high accuracy of the \tool: zero false positives were found, indicating that all detected violations matched the reported issues. 

\begin{table}[t]
\rowcolors{1}{}{gray!10}
\caption{Manual Validation Results of the detected violations}
\centering
    \resizebox{1\linewidth}{!}{%
    \renewcommand{\arraystretch}{1}
    \setlength{\tabcolsep}{4pt}
\begin{tabular}{lcc}
\toprule
\textbf{Violation Category} & \textbf{\# Violations Checked} & \textbf{False Positives} \\
\midrule
Class Names & 36 & None \\

Method Names & 42 & None \\

Variable Names & 39 & None \\

Package Names & 31 & None \\

Javadocs Class & 43 & None \\

Javadoc Method & 44 & None \\

Javadoc Constructor & 45 & None \\

Javadoc Formatting & 43 & None \\

Private Instances & 36 & None \\

Useless & 40 & None \\

String Concatenation & 31 & None \\

Missing Override & 31 & None \\

Finalize Override & 31 & None \\

Unqualified Static Access & 33 & None \\

Empty Catch Block & 35 & None \\
\bottomrule
\end{tabular}
}
\label{tab:violation-checking}
\end{table}

\section{Experimental Results}

\subsection{RQ1 - Violation Prevalence}
RQ1: \emph{How prevalent are different code style and programming practice violations in open-source \java projects?}

\medskip
\noindent
\textbf{Absolute number of violations.}
Table~\ref{tab:violation-stats} provides statistical insights into the absolute numbers of violations detected across the analyzed 1,036 \java repositories. The data highlights significant variability across different categories of code style and programming practice violations. All categories also had minimum values of 0, meaning every category had at least 1 repository with no violations.

\smallskip
Notably, Javadoc-related violations occur most frequently. For instance, \emph{Javadoc Formatting} violations have the highest average (380.18) and the maximum number (6,213), indicating widespread inconsistencies in documenting code. Similarly, \emph{Javadoc Method} violations also have a high average (303.26), reflecting substantial issues in method documentation across repositories.

\smallskip
Violations related to naming conventions, such as \emph{Variable Names}, \emph{Method Names}, and \emph{Class Names}, also appear frequently. \emph{Variable Names} show the highest average (88.34) among naming conventions, suggesting common issues with variable naming standards across repositories.

\begin{figure*}[t]
\centering
\includegraphics[width=0.8\linewidth]{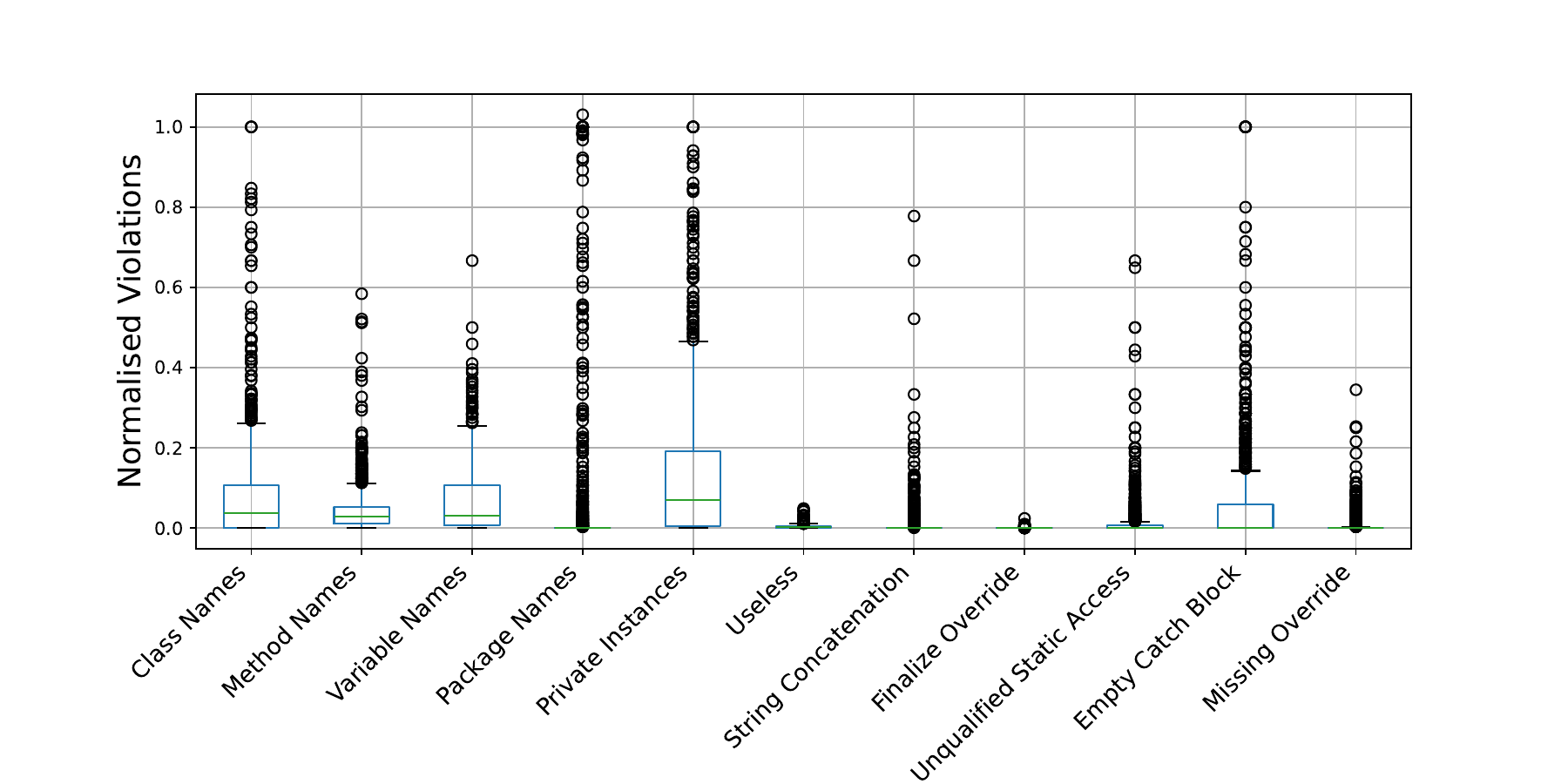}
\vspace{-2mm}
\caption{Distribution of normalized violation counts across categories (RQ1) }
\label{fig:all-categories}
\end{figure*}

\begin{table}[t]
\rowcolors{1}{}{gray!10}
    \centering
    \caption{Absolute Violation Statistics Across the 1,036 repos (RQ1)}
    \renewcommand{\arraystretch}{1} % Fix array stretch issue
    \setlength{\tabcolsep}{8pt}  % Adjust column spacing

    \resizebox{1\linewidth}{!}{ % Resize only the tabular part
    \begin{tabular}{lrrrr}
        \toprule
        \textbf{Violation Category} & \textbf{Min} & \textbf{Max} & \textbf{Avg} & \textbf{Median} \\
        \midrule
        Class Names & 0 & 231 & 9.45 & 3.0 \\
        Method Names & 0 & 385 & 26.45 & 9.0 \\
        Variable Names & 0 & 2,514 & 88.34 & 28.0 \\
        Package Names & 0 & 308 & 4.30 & 0.0 \\
        Javadoc Class & 0 & 768 & 56.53 & 25.0 \\
        Javadoc Method & 0 & 5,159 & 303.26 & 144.0 \\
        Javadoc Constructor & 0 & 516 & 55.91 & 29.0 \\
        Javadoc Formatting & 0 & 6,213 & 380.18 & 124.0 \\
        Private Instances & 0 & 1,025 & 30.02 & 8.0 \\
        Useless & 0 & 920 & 36.09 & 13.5 \\
        String Concatenation & 0 & 88 & 0.56 & 0.0 \\
        Finalize Override & 0 & 15 & 0.09 & 0.0 \\
        Unqualified Static Access & 0 & 135 & 2.76 & 0.0 \\
        Empty Catch Block & 0 & 215 & 2.50 & 0.0 \\
        Missing Override & 0 & 321 & 3.17 & 0.0 \\
        \bottomrule
    \end{tabular}
    }
    
    \label{tab:violation-stats}
\end{table}

\smallskip
Programming practice violations generally show lower averages compared to code style violations. For example, \emph{String Concatenation} (0.56), \emph{Finalize Override} (0.09), and \emph{Empty Catch Block} (2.50) appear far less frequently. The particularly low frequency of \emph{String Concatenation} violations suggests that developers of popular \java repositories on \gh are aware of the (very serious) performance issues of string concatenation in loops and actively avoid it, or that there tends to be few scenarios of string concatenation in loops.

\medskip
\noindent
\textbf{Normalized number of violations.} It is important to highlight that \tool calculates violations both in absolute values and as normalized metrics relative to project size~\cite{gradestyle}. Specifically, violations are normalized based on the total number of relevant constructs (e.g., variables, classes, methods). For example, \emph{VariableNames} violations are adjusted according to the total number of declared variables in a repository. This normalization ensures fairness, preventing larger projects from being disproportionately penalized simply due to their size. We argue that normalized values provide a more meaningful metric for comparing code quality across repositories of different sizes. Indeed, as previously reported the \java LOC of the projects ranges from 29 to $\sim$9M.

\smallskip
Figure~\ref{fig:all-categories} presents the distribution of all non-Javadoc-related violations in their normalized form.

\smallskip
All categories exhibit a top-skewed distribution, where most repositories have relatively few normalized violations, while a smaller subset of repositories show significantly high violation counts. This indicates that while the majority of repositories adhere to coding standards, a fraction of them exhibit poor compliance with code style and programming best practices.

\smallskip
The results largely align with the absolute violation counts, with one key exception: the \emph{Empty Catch Block} category shows a notably higher violation rate in the normalized distribution. While its median remains at 0 (indicating that at least half of the repositories have no violations of this type), the upper quartile reaches 0.0588, much higher than other programming practice violations. In contrast, other programming violations such as \emph{Unqualified Static Access} (0.0066), \emph{Missing Override} (0.0014), and \emph{Finalize Override} (0.0) have considerably lower upper quartile values. This means that the worst 25\% of repositories contain empty catch blocks in over 5.8\% of their catch statements. Notably, one repository had a normalized violation rate of 1, indicating that every single catch block in the project was empty.

\begin{table*}[ht]
\rowcolors{1}{}{gray!10}
\centering
\caption{Normalized violation thresholds by category and type (RQ2)}
\label{tab:thresholds}
\renewcommand{\arraystretch}{1} % Adjust row spacing
\setlength{\tabcolsep}{6pt}  % Adjust column spacing

\resizebox{\textwidth}{!}{%
\begin{tabular}{@{} l l c c c c || c  c c c c c c @{}}
    \toprule
    \textbf{Category} & \textbf{Type} & \textbf{25\%} & \textbf{20\%} & \textbf{15\%} & \textbf{10\%} &  \textbf{5\%} & \textbf{4\%} & \textbf{3\%} & \textbf{2\%} & \textbf{1\%} & \textbf{0\%} \\
    \midrule
   Class Names & Code Style & 93.34 & 89.77 & 83.69 & 73.17 & 55.41 & 50.77 & 44.98 & 39.09 & 34.27 & 30.98 \\
    Method Names & Code Style & 98.94 & 98.36 & 96.24 & 92.37 & 73.07 & 62.64 & 52.12 & 35.04 & 22.10 & 13.03 \\
   Variable Names & Code Style & 96.53 & 93.05 & 85.91 & 73.17 & 59.07 & 55.21 & 49.42 & 41.12 & 28.09 & 9.65 \\
    Package Names & Code Style & 93.53 & 92.95 & 92.28 & 91.51 & 89.19 & 88.42 & 87.16 & 85.42 & 83.49 & 80.98 \\
    \midrule
    %Private Instances & Programming Practice & 81.56 & 76.74 & 68.72 & 57.33 & 44.79 & 41.60 & 37.84 & 32.14 & 26.64 & 22.30 \\
    %\textbf{Useless} & Programming Practice & 100.0 & 100.0 & 100.0 & 100.0 & 100.0 & 99.61 & 99.32 & 98.74 & 92.08 & 7.43 \\
    Finalize Override & Programming Practice & 100.0 & 100.0 & 100.0 & 100.0 & 100.0 & 100.0 & 100.0 & 99.90 & 99.90 & 95.37 \\
    Unqualified Static Access & Programming Practice & 99.03 & 98.84 & 97.97 & 96.72 & 93.15 & 91.70 & 89.09 & 86.10 & 79.15 & 67.18 \\
   Empty Catch Block & Programming Practice & 94.21 & 91.89 & 88.71 & 83.20 & 73.07 & 70.75 & 66.80 & 64.67 & 61.97 & 60.33 \\
    Missing Override & Programming Practice & 99.81 & 99.61 & 99.42 & 99.03 & 97.01 & 96.24 & 95.56 & 94.02 & 89.48 & 71.14 \\
    \bottomrule
\end{tabular}
} % End resizebox

\end{table*}

\smallskip
Despite this, programming practice violations remain less frequent than code style violations, suggesting better adherence to best practices in \java development. However, two notable exceptions stand out: \emph{Private Instances} and \emph{Empty Catch Blocks}. These two categories show violation rates comparable to the more frequently occurring code style violations.

\begin{figure}[t]
\centering
\includegraphics[width=1\linewidth]{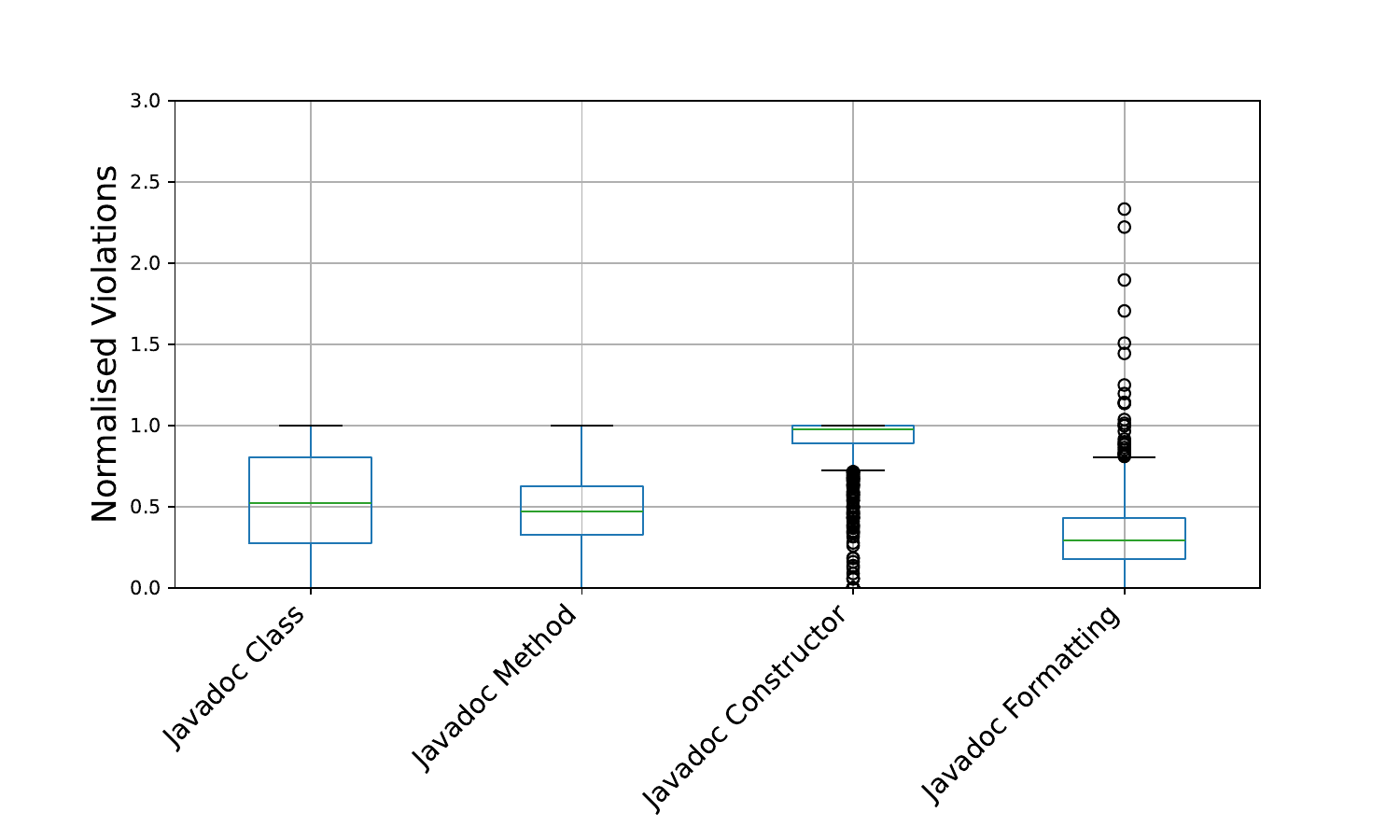}
\vspace{-8mm}
\caption{Prevalence of Javadoc Violations (RQ1)}
\label{fig:Javadoc-violations}
\end{figure}

\smallskip
Javadoc-related violations stand apart due to their significantly higher normalized values, as shown in Figure~\ref{fig:Javadoc-violations}. The medians of Javadoc categories are 0.52, 0.47, 0.97, and 0.29, which are much higher than those of other categories. Notably, all Javadoc categories except \emph{Javadoc Formatting} have a maximum normalized value of 1, meaning that some repositories lack Javadoc documentation entirely. \emph{Javadoc Formatting} violations, however, exceed this threshold, with some repositories having normalized values greater than 1. This occurs because a single Javadoc comment can have multiple associated violations. One extreme outlier had a Javadoc Formatting violation rate of 6, meaning that, on average, each Javadoc comment contained six violations—the maximum possible number. These findings suggest that Javadoc compliance is a major issue in open-source \java repositories.

\begin{custombox}{RQ1 in Summary}
Our analysis of 1,036 open-source \java repositories reveals that Javadoc violations are the most frequent, while naming and programming practice violations occur less often. \emph{Empty Catch Blocks} and \emph{Private Instances} stand out as key areas for improvement.
\end{custombox}

\subsection{RQ2 - Claimed vs. Actual Code Style Adherence}
RQ2: \emph{Do repositories that explicitly claim to follow a code style guide exhibit better adherence to coding standards than those that do not?}

\medskip
RQ2 investigates whether repositories that claim adherence to a specific code style, particularly \textsc{Google}’s \java Style Guide~\cite{google}, exhibit higher compliance with coding standards than those that do not. Additionally, it examines whether repositories that strictly follow a code style explicitly document their adherence or if it remains implicit.

\smallskip
To identify repositories claiming adherence, we performed keyword searches on all markdown (\texttt{.md}) files located at the project's root using regular expressions to detect references to code style. We used two sets of regular expressions; one searches for general code-style references, such as ``code style'' and ``coding standards''. The other is more specific, and searches for \textsc{Google} \java Style mentions, such as "\textsc{Google} Java Style" or the official URL of the guide. We included all root markdown (\texttt{.md}) files to ensure we covered common files such as \texttt{README.md} and \texttt{CONTRIBUTING.md} where code-style and best-practice expectations are typically conveyed to contributors. We also checked for the presence of PMD or \checkstyle configuration files, as these are the two most widely used static analysis tools for enforcing \java coding standards. 
While this automated approach cannot guarantee complete precision, it represents a best-effort solution. Indeed, manually analyzing all 1,036 repositories would be too costly.

\smallskip
The repositories were classified into three categories:

\begin{itemize}
    \item \textbf{Mention of Code Style} (140 repositories): A root documentation file contains general references to code style, such as ``code style'' or ``code-style''.
    \item \textbf{Explicit Mention of \textsc{Google}’s \java Style Guide} (16 repositories): The repository explicitly references \textsc{Google}’s \java Style Guide, includes a link to its official documentation in a root markdown (\texttt{.md}) file, or contains configuration files for automated \textsc{Google} style enforcement (e.g., \texttt{checkstyle.xml}).
    \item \textbf{No Mention} (880 repositories): The repository does not explicitly reference any coding standard.
\end{itemize}

Next, we analyzed the prevalence of violations by defining a threshold to determine when a repository can be considered adherent. Table~\ref{tab:thresholds} presents the percentage of repositories falling below specific normalized violation thresholds. For instance, 94.21\% of repositories have normalized \emph{Empty Catch Block} violations below 0.25. Note that Table~\ref{tab:thresholds} only includes violations explicitly defined in the \textsc{Google} \java Style Guide (see the \textsc{Google} column in Table~\ref{tab:violations-tools}). As a result, we excluded \emph{Private Instances}, \emph{Useless}, and \emph{String Concatenation}, since these are not part of the official style guide. Additionally, we excluded Javadoc-related violations because RQ1 shows high frequency, making them less informative for characterizing code style adherence. 

\smallskip
Based on this analysis, we set a threshold at 5\% (0.05 normalized violations) to classify repositories as adherent, as the number of repositories passing this threshold stabilizes beyond this point. 
Setting the threshold to 0 would be unrealistic, as it assumes that repositories are entirely free of violations, which is rarely the case in practice. Instead, a repository is considered adherent to a code style category if its normalized violations fall below the chosen threshold.

\begin{figure}[t]
\centering
\includegraphics[width=0.82\linewidth]{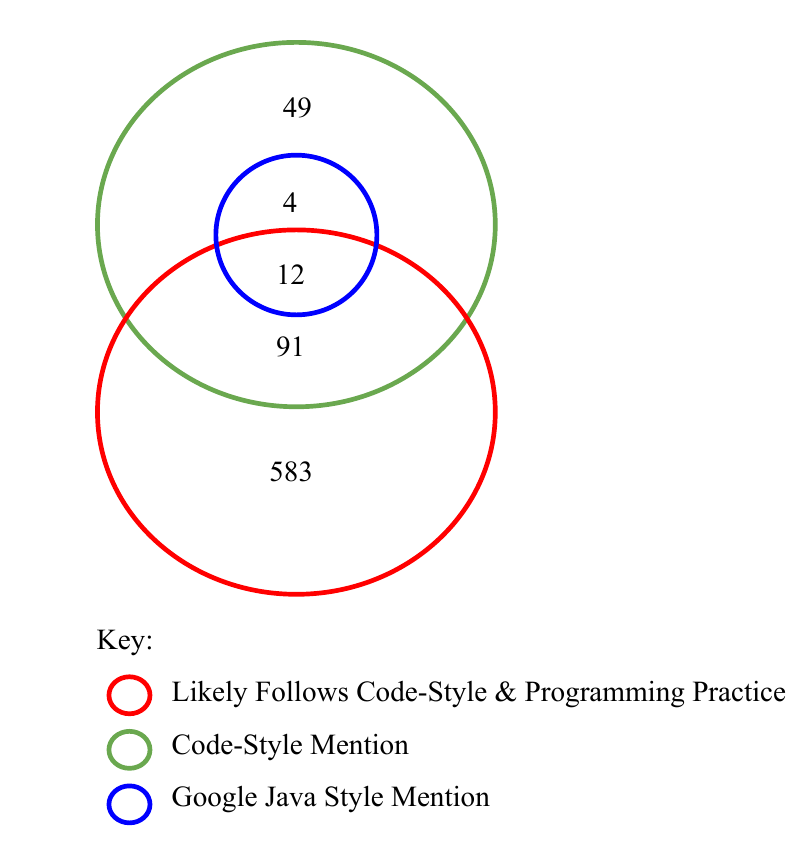}
\vspace{-4mm}
\caption{Claimed Adherence vs. Actual Adherence (RQ2)}
\label{fig:venn}
\end{figure}

\smallskip
To assess whether claimed adherence aligns with actual adherence, we analyzed the distribution of programming practice and code style violations across repositories. Figure~\ref{fig:venn} presents a Venn diagram illustrating how adherence rates compare to repositories' self-reported compliance.

\smallskip
The Venn diagram reveals that 12 out of 16 repositories (75\%) that explicitly claim adherence to \textsc{Google}’s \java Style Guide meet the 5\% threshold for programming practice categories. This adherence rate is higher than the 65\% observed among repositories that mention code style and the 66\% among those with no explicit reference to coding standards.

\begin{custombox}{RQ2 in Summary}
Repositories that explicitly reference \textsc{Google}’s \java Style Guide demonstrate the highest adherence rates to code style and programming practices, with 75\% meeting the 5\% compliance threshold. However, adherence remains relatively high (65--66\%) even among repositories that do not explicitly claim compliance, suggesting that many projects follow best practices implicitly.
\end{custombox}

\subsection{RQ3 - Evolution of Code Style and Best Programming Practices}
RQ3: \emph{How does code style and programming practice adherence evolve over time in actively maintained and mature \java repositories?}

\begin{figure}[t]
\centering
\includegraphics[width=1.0\linewidth]{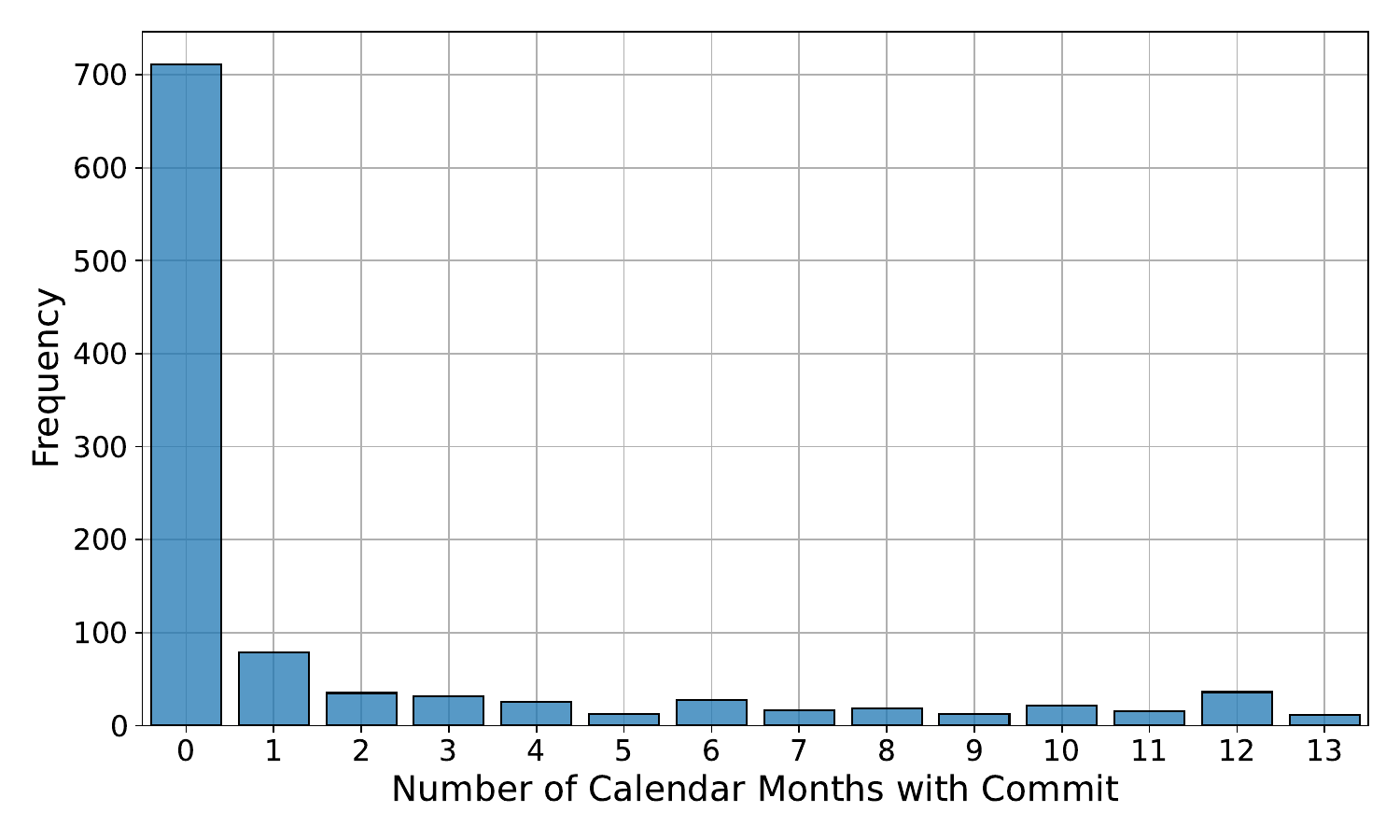}
\vspace{-8mm}
\caption{Monthly Commit Activity over 1 Year (RQ3) }
\label{fig:monthly-commit-distribution}
\end{figure}

\smallskip
To answer RQ3, we examined the 1,036 repositories to identify those that are both mature and actively maintained. We analyzed commit histories over multiple time intervals, collecting timestamps of all commits made in the past year and counting how many intervals contained at least one commit. We then examined the results to identify repositories with consistent activity. Figure~\ref{fig:monthly-commit-distribution} illustrates the distribution of monthly commits over a one-year period for the 1,036 repositories.

\smallskip
From this analysis, we found that 43 repositories have at least one commit every calendar month in the past year. Active maintenance was a crucial selection criterion, as regularly updated repositories with frequent commits provide a more reliable dataset for tracking code evolution over time. To further refine the selection, we filtered for mature repositories: those that are at least 36 months old. This criterion ensures that projects have had enough time to refine their coding standards. After applying this filter, 41 repositories were retained.

\begin{figure}
\centering
\includegraphics[width=1\linewidth]{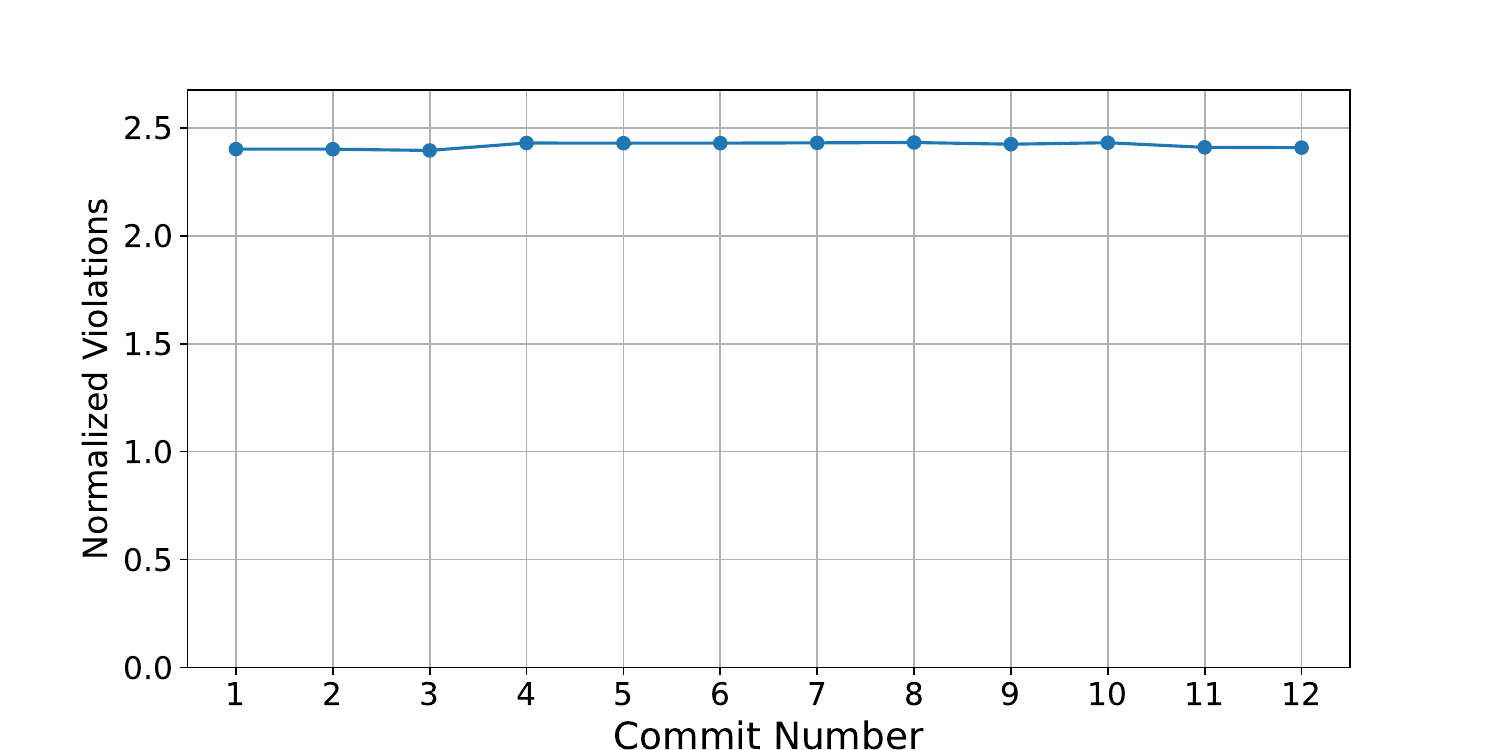}
   \vspace{-6mm}
\caption{Variation of Total Normalized Violations with Time (RQ3)}
\label{fig:total-normalised-score-trend}
\end{figure}

\smallskip
For each of these 41 repositories, we sampled 12 commits from the past year, selecting one commit per calendar month. The commit chosen for each month was the one closest to the middle of the month. For example, if a repository had commits on January 2nd, 10th, and 27th, the commit on January 10th was selected as it is closest to the median day of the month (15th). We considered only the last year to avoid analyzing repositories too early in their lifecycle, as early-stage projects may exhibit immature coding practices or function as proof-of-concept projects rather than mature projects.

\smallskip
To prevent selecting commits that were too close in time, we computed the smallest differences between commit dates and manually inspected them. The analysis confirmed that no two commits from the same repository were closer than 10 days apart, ensuring a well-distributed selection.

\begin{figure*}[t]
    \centering
    \begin{minipage}[t]{0.49\linewidth}
        \centering
        \includegraphics[width=\linewidth]{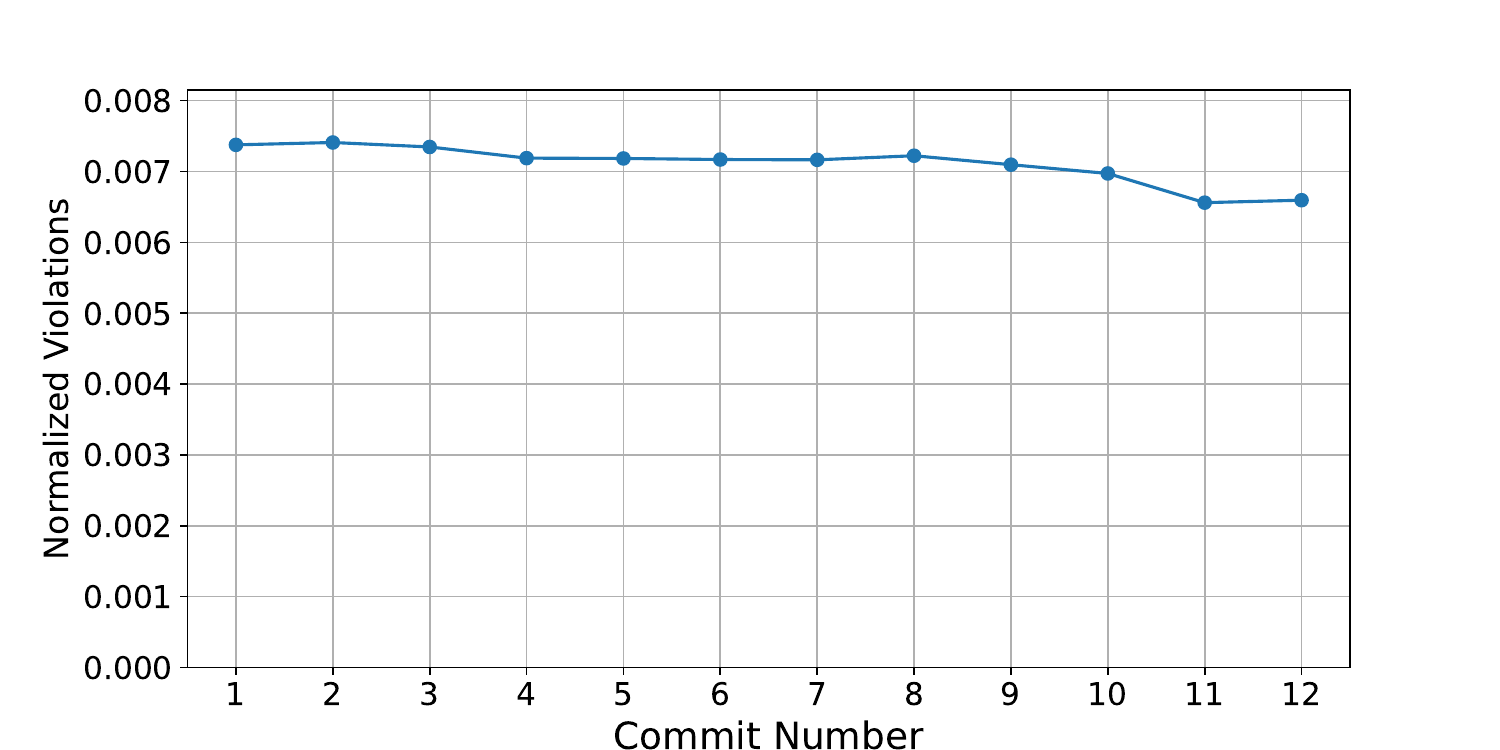}
        \vspace{-6mm}
        \caption{Trend of Programming Practice Normalized Violations (RQ3)}
        \label{fig:programming-pratice-trend}
    \end{minipage}
    \hfill
    \begin{minipage}[t]{0.49\linewidth}
        \centering
        \includegraphics[width=\linewidth]{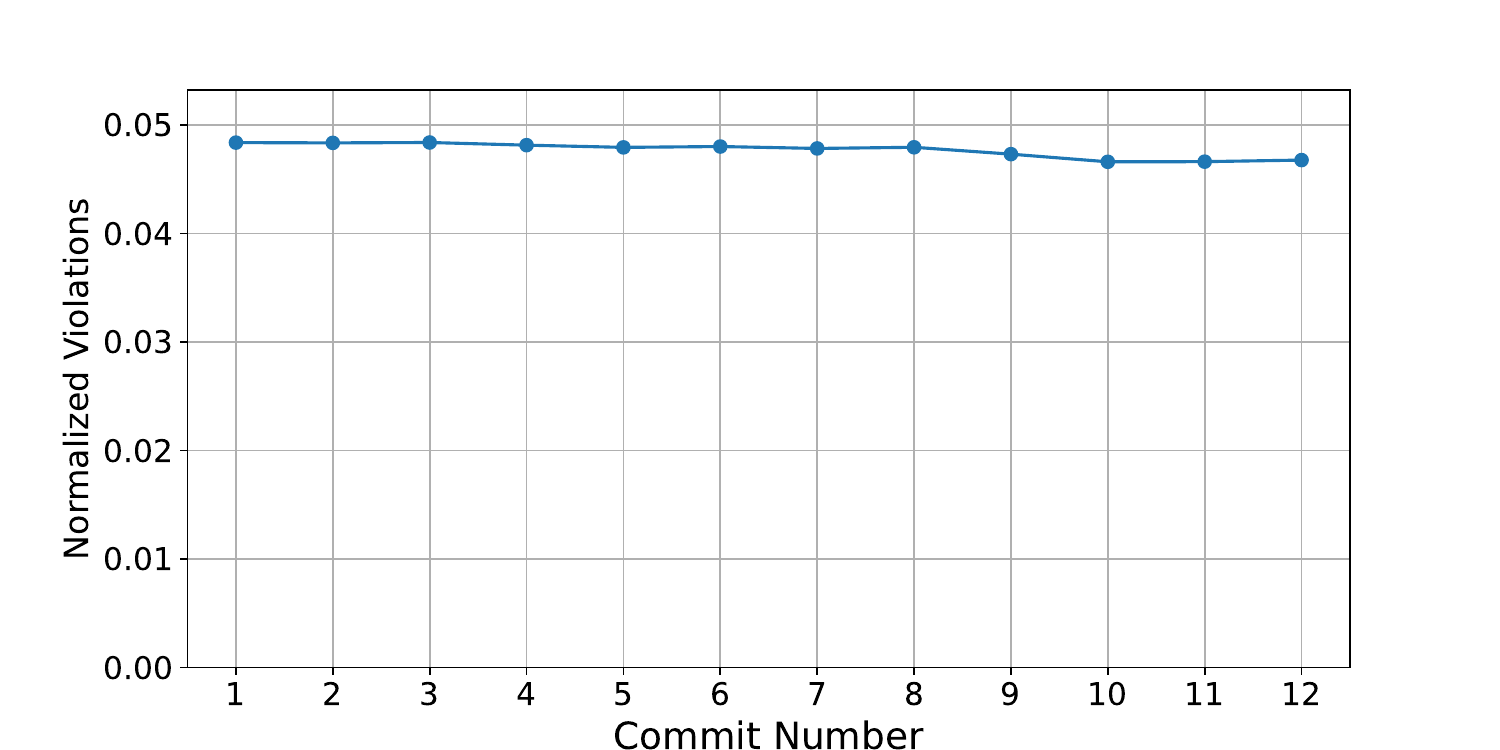}
        \vspace{-6mm}
        \caption{Trend of Class Names Normalized Violations (RQ3)}
        \label{fig:class-names-trend}
    \end{minipage}
    \vspace{-2mm}
\end{figure*}
\begin{figure*}[t]
    \centering
    \begin{minipage}[t]{0.49\linewidth}
        \centering
        \includegraphics[width=\linewidth]{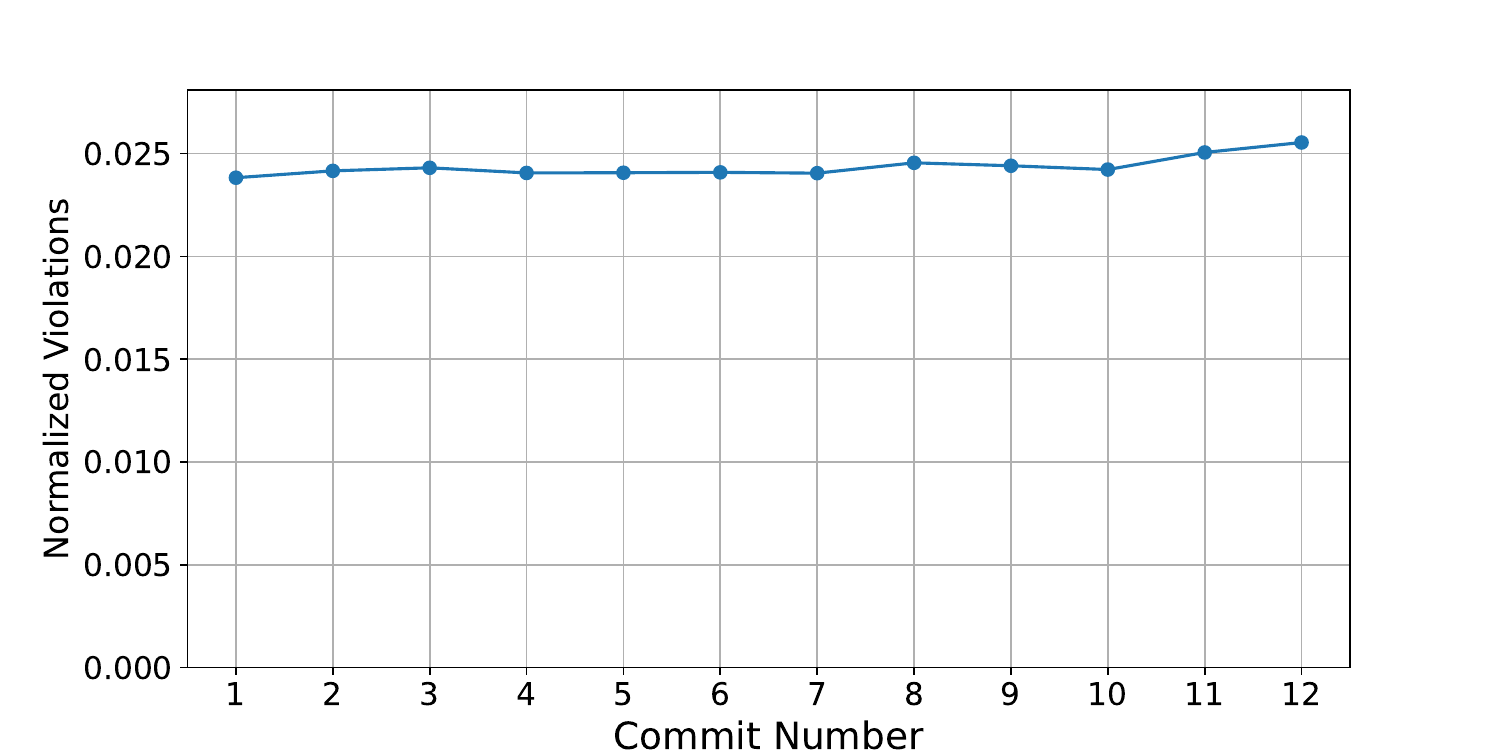}
        \vspace{-6mm}
        \caption{Trend of Variable Naming Normalized Violations (RQ3)}
        \label{fig:variable-naming-trend}
    \end{minipage}
    \hfill
    \begin{minipage}[t]{0.49\linewidth}
        \centering
        \includegraphics[width=\linewidth]{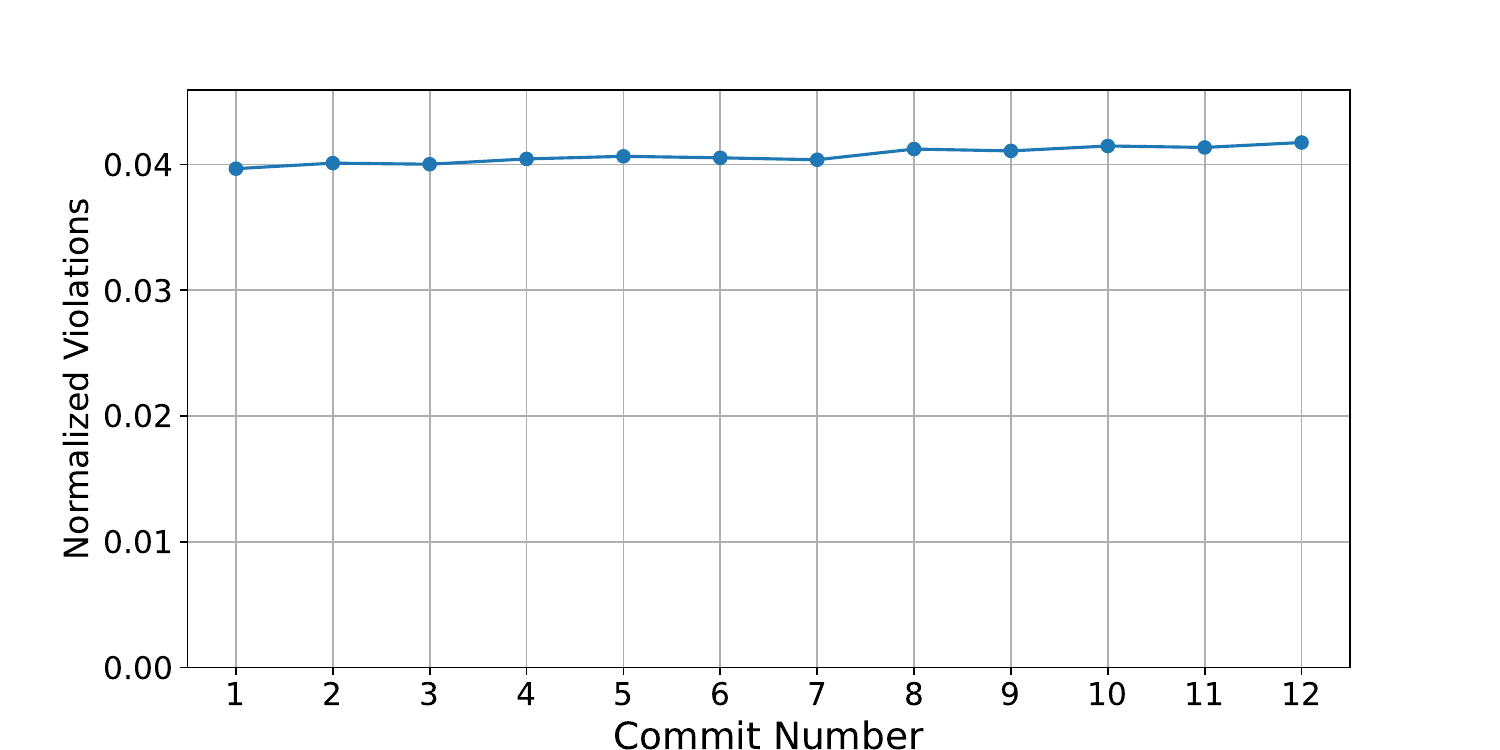}
        \vspace{-6mm}
        \caption{Trend of Method Naming Normalized Violations (RQ3)}
        \label{fig:method-naming-trend}
    \end{minipage}
\end{figure*}

\smallskip
Finally, we ran the tool on each commit across all selected repositories. Each repository was analyzed 12 times, starting from the oldest commit (February 2024) to the most recent (January 2025). For each commit, the repository was checked out to the corresponding commit hash, restoring the codebase to its exact state at that point in time.  Figure~\ref{fig:total-normalised-score-trend} shows the average normalized score for each commit across all 41 repositories. Overall, the total normalized score remains relatively stable over the 12-month period, with a slight increase of 0.0068 from February 2024 to January 2025. The variation is minimal, with a maximum fluctuation of 0.037, suggesting that code style and programming practice adherence remains largely unchanged.

\smallskip
Breaking down the results by category, we observe:
\begin{itemize}
\item \textbf{Programming practice violations} (Figure~\ref{fig:programming-pratice-trend}) show a slight downward trend, decreasing from 0.0074 to 0.0066 over the 12 months, suggesting gradual improvement in programming best practices.
\item \textbf{Class name violations} (Figure~\ref{fig:class-names-trend}) also decrease, with a small drop of 0.0014
\item \textbf{Variable and method naming violations} (Figures~\ref{fig:variable-naming-trend} and~\ref{fig:method-naming-trend}) exhibit a slight upward trend, indicating a gradual decline in adherence to naming conventions.
\end{itemize}

Other categories showed no clear trends.

\begin{custombox}{RQ3 in Summary}
Our analysis of actively maintained and mature repositories suggests that code style programming practice adherence remains relatively stable. Notably, class name and programming practices violations slightly decrease, whereas variable and method naming violations slightly increase.
\end{custombox}

\subsection{RQ4 - Common Orderings of Class Elements}
RQ4: \emph{Is there a common ordering of \java elements in a class that is widespread in open-source repositories?}
\label{sec:o}

\smallskip
The order of elements inside a class significantly impacts code \emph{readability}\cite{google}. While \textsc{Google}’s \java Style Guide acknowledges its importance, it does not enforce a specific ordering\cite{google}. Although we did not include ordering violations in our primary study, we want to know if a most common one exists.  To achieve this, we analyzed the 1,036 repositories, evaluating different class element arrangements, including fields, constructors, instance methods, and static methods. This check is not implemented in \checkstyle (see Section 3.4.2 in~\cite{checkstyleGoogle}), presumably due to the lack of general agreement on a standard ordering. In contrast, it is supported by \tool. We extended \tool by adding configurable ordering checks and evaluated four commonly used conventions.

\smallskip
\mbox{\textbf{Ordering 1:}} Inner Classes, Static Fields, Static Methods, Instance Fields, Constructors, Instance Methods (original order used in the \tool paper~\cite{gradestyle}).

\smallskip
\mbox{\textbf{Ordering 2:}} Static Fields, Static Methods, Instance Fields, Constructors, Instance Methods, Inner Classes.

\smallskip 
\mbox{\textbf{Ordering 3:}} Static Fields, Static Methods, Instance Fields, Instance Methods, Constructors, Inner Classes

\smallskip
\mbox{\textbf{Ordering 4:}} Instance Fields, Constructors, Instance Methods, Static Fields, Static Methods, Inner Classes.

\begin{figure}[t]
\centering
\includegraphics[width=1\linewidth]{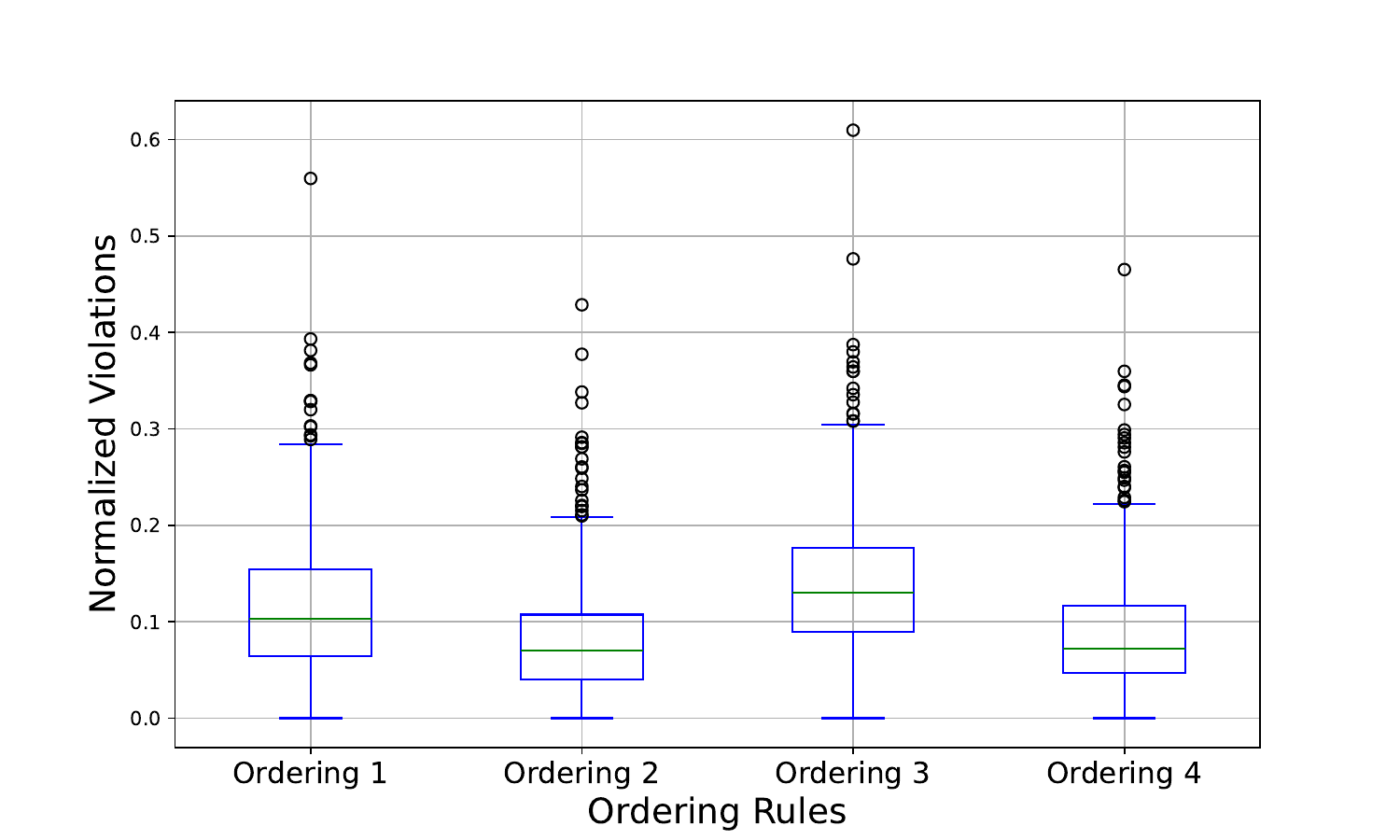}
\vspace{-4mm}
\caption{Comparison of Ordering Rules (RQ4)}
\label{fig:ordering-comparison}
\end{figure}

\smallskip
Figure~\ref{fig:ordering-comparison} presents the normalized scores for each ordering across the 1,036 repositories.

\smallskip
The results show that \textbf{Orderings 2 and 4} are the most commonly followed, with medians of 0.070 and 0.071, respectively—lower than Orderings 1 and 3 (0.10 and 0.13). This suggests that developers prefer structuring classes with static members grouped together or prioritizing instance-related elements first. A key actionable insight from our study is that open-source repositories should adopt either Ordering 2 or 4.

\begin{custombox}{RQ4 in Summary}
Our analysis revealed two popular orderings of \java Class elements being used in the repositories. One places static fields and methods at the beginning of the class, while the other places them at the end. We also found that no one ordering was universal among the repositories.
\end{custombox}

\section{Discussion}
This section provides a discussion of the key findings, analyzing their implications for code style adherence, programming best practices, and the broader impact on open-source software development

\subsection{Code Style Adoption}
Our findings highlight a need for better adherence to code style conventions in open-source \java projects. Figure~\ref{fig:all-categories} shows high normalized scores for key categories such as class names, package names, variable names, and private instances, suggesting these areas require more attention from developers.

Conversely, normalized scores for programming practice and \emph{useless code} categories indicate that these practices are largely followed. The primary exception is the \emph{Empty Catch Block} category, which exhibits relatively high violation rates. This is concerning as empty catch blocks hide errors and make debugging harder, which is why it is part of the \textsc{Google} \java Style guide. Therefore more attention needs to be paid to properly make use of catch statements.

\subsection{Explicit vs. Implicit Adherence}
Interestingly, repositories that explicitly reference \textsc{Google}’s \java Style Guide~\cite{google} exhibit the best adherence to programming practice violations. However, only 16 repositories made such claims, highlighting a lack of formal adoption. Encouraging more projects to reference and enforce established style guides could improve overall code quality.

\smallskip
Our results may also encourage owners of open-source \java projects to actively push good practices more than they do currently. Further standardizing of \java programming practices could also help to push this, for example by seeing common agreement on the ordering of elements within a \java class. Furthermore, enforcement of more novel violations should be overseen by managers, such as missing \textit{@Override} annotations. These are especially important due to the annotation helping other developers to understand where subclasses are used. We hope that by releasing our tool with enhanced violation checkers, we can contribute to improving code style adherence within the \java community.

\section{Threats to Validity}

\smallskip \noindent
\textbf{Tool Effectiveness.} The manual validation step focused only on false positives (i.e., incorrectly flagged violations) but did not systematically evaluate false negatives (i.e., undetected violations). If present, false negatives would indicate our results under-report the prevalence of violations, meaning the violations analysed in this study may be even more widespread. Despite these limitations, the existence of false negatives can only increase the number of true violations, not decrease. This means that our tool can still confidently serve as a lower limit for the number of violations, as we found zero false positives in our manual analysis. While we conducted limited testing for false negatives when extending the tool, a thorough false negative evaluation remains an important future work.

\smallskip \noindent
\textbf{Time Frame for Evolution.} Our study analyzed code evolution over the last one-year period, which may not be sufficient to capture long-term trends. Future research could extend this analysis to cover longer periods (e.g., 5–10 years) to better understand how code style adherence evolves in mature projects. Additionally, we sampled one commit per month, a granularity that may be too coarse for highly active repositories. This approach may have missed sudden bursts of code style activity or finer-grained changes in coding behavior. Future studies could employ more refined sampling strategies (e.g., sliding windows or multiple commits per month) to uncover subtler temporal dynamics in coding practices.

\smallskip \noindent
\textbf{Selection Bias.} The study focused on 1,036 of the most starred repositories on \gh. While this selection ensures popularity and relevance, it may introduce bias, as highly starred repositories may not be representative of the broader open-source \java ecosystem, including smaller or less-maintained projects. Exploring repositories with fewer stars remains an important direction for future work to achieve a more comprehensive understanding of the ecosystem.

\smallskip \noindent
\textbf{Threshold Sensitivity.} The results are influenced by the chosen 5\% violation threshold, and a different threshold could lead to varying conclusions about adherence rates. We mitigate this threat by employing a data-driven approach to determine this threshold (See Table~\ref{tab:thresholds}). We believe it effectively distinguishes adherence from non-adherence.

\section{Related Work}

To the best of our knowledge, our study is the first of its kind in terms of scale, analyzing a large number of repositories and detecting coding violations that have not been examined in previous research. Moreover, we address novel research questions and provide new insights into code style adherence, its evolution over time, and the gap between claimed and actual adherence in open-source \java projects. We now discuss the most related work.

\smallskip
Coding style conventions and best programming practices are well-established principles that enhance software readability and maintainability~\cite{elish2002adherence}. Many programming languages offer official style guides, such as \textsc{Python}’s PEP8 and \textsc{Google}’s coding standards. However, studies on open-source projects suggest that adherence is inconsistent~\cite{DBLP:conf/icse/MotwaniB23a}.  Beller et al.~\cite{Beller2016} conducted a study of static analysis tools (including linters for style) across popular projects in \java, \textsc{JavaScript}, \textsc{Python}, and \textsc{Ruby}~\cite{Beller2016}. They found that roughly half of the studied projects employ at least one automated static analysis tool to enforce best practices, but usually in an ad-hoc manner. Notably, coding style rules are not always strictly enforced—many teams use these tools to flag issues but do not mandate a zero-violation policy. Similarly, Boogerd et al.~\cite{Boogerd2008} observed that although style checkers raise a significant portion of build warnings, they rarely cause build failures, suggesting lenient enforcement. Our study builds on these insights by examining adherence trends specifically in \java and investigating whether explicit claims of compliance translate to actual adherence (RQ2).

\smallskip
The \tool study~\cite{gradestyle} investigated coding style in student repositories, showing improvement in adherence after receiving automated feedback. Brown et al.~\cite{brown2022novice} similarly studied novice programmers, observing slight increases in Javadoc use over time. While these works analyze controlled learning environments, our study extends this approach to real-world open-source projects.

\smallskip
Jonsson et al.\cite{jonsson2023opensource} examined adherence to \textsc{Oracle}’s \java conventions in a limited set of open-source projects, using PMD and \checkstyle for violation detection. Similarly, Elish and Offutt~\cite{elish2002adherence} analyzed 100 \java projects and found that only 4\% of classes fully adhered to coding conventions. Both of these studies mainly focus on naming, commenting, and structure. Among naming related violations, Elish and Offutt found that 28 out of their 100 investigated \java classes had field naming violations, and 18 out of 100 classes had method naming violations. While it is hard to directly compare our results due to a different normalisation approach, these results line up roughly with those of our study, with most field and method names following convention. Our results also agree that field violations are generally more common than method violations. However, Elish and Offutt found no violations of class naming, a strong contrast to our findings, where class naming violations were the most frequent of the naming violations. This suggests that our NLP-based tool can detect naming violations missed by tools limited to camelCase checks.

\smallskip
While our study includes some of these aspects, we also consider violations that were not covered, or only partially addressed. For example, we check whether class names are nouns/noun phrases and method names are verbs/verb phrases, something not checked in either study. Our study also differs by including novel violation checks not covered by previous research, such as \emph{Missing Override} and \emph{Unqualified Static Access}. We also use a significantly larger dataset of 1,036 repositories and cover a broader range of violations, providing a more comprehensive view of modern \java coding practices.

\section{Conclusions and futue work}

We provided a comprehensive and novel analysis of adherence of code style and best programming practices in 1,036 open-source \java projects on \gh. Our findings reveal that Javadoc-related violations are the most prevalent. We also observe that there is still room for improvement in other categories, such as naming conventions, private instances and other programming practices, such as empty catch blocks. Repositories that explicitly claim adherence to \textsc{Google}'s \java Style Guide demonstrate noticeably better compliance. This highlights the importance of formalizing coding standards and expectations in open-source projects. By releasing the updated \tool~\cite{gradestyletool} and all experimental data~\cite{data}, we aim to foster further research and encourage the open-source community to prioritize consistent coding standards. %Such work contributes to ongoing effort to improve software quality in open-source \java projects. 
We submitted 14 pull requests fixing violations detected by our tool. One has been confirmed\footnote{\url{https://github.com/x-ream/sqli/pull/61}}, and another merged\footnote{\url{https://github.com/Aliucord/Aliucord/pull/521}}.

\smallskip
While this study provides valuable insights into code style adherence and programming practice violations in open-source \java projects, several areas remain open for future research. We now highlight the three most promising ones.

\smallskip
First, our study examined code style evolution over a one-year period. Future research could extend this analysis to longer time frames to better understand long-term trends and the factors influencing adherence over time.

\smallskip
Second, while our study focuses on the most starred \java open-source repositories on \gh, future work could explore the adoption of code style and programming practices in repositories with a wider range of star counts. This is important because highly starred repositories may not be fully representative of the open-source ecosystem, where smaller or less popular projects may exhibit different adherence patterns and challenges in maintaining coding standards.

\smallskip
Third, although our dataset likely includes a variety of project types (such as libraries, frameworks, and applications) we did not explicitly classify them. Differentiating between these categories could uncover domain-specific adherence patterns and help identify which types of projects are more or less compliant with established coding standards. This represents an interesting direction for future work. 

% Open-source development plays a crucial role in modern software engineering~\cite{spinellis2004opensource}, and maintaining consistent code style is essential for readability and maintainability~\cite{tornhill2022codered}. Our work contributes to this area by providing a large-scale analysis of adherence patterns, highlighting areas where enforcement can be improved, and tracking changes over time.

\bibliographystyle{IEEEtran}  
\bibliography{bib}  
\end{document}